\documentclass[preprint,showpacs,preprintnumbers,amsmath,amssymb,superscriptaddress]{revtex4-2}
\usepackage{graphicx}
\usepackage{amsmath}
\usepackage{amssymb}
\usepackage{epsfig}
\usepackage{amsthm}
\usepackage{color}
\usepackage{bm}
\def\be{\begin{equation}}
\def\ee{\end{equation}}
\def\arr{\begin{array}{rll}}
\def\ea{\end{array}}
\def\bea{\begin{eqnarray}}
\def\eea{\end{eqnarray}}

\begin{document}
\title{Quantum Chromodynamics of the Nucleon in Terms of Complex Probabilistic Processes}

\author{Ashot S. Gevorkyan}
\affiliation{Institute for Informatics and Automation Problems NAS of RA,
1, P. Sevak str., Yerevan, 0014, Republic of Armenia}
\affiliation{Institute of Chemical Physics,  NAS of RA, 5/2, P. Sevak str., Yerevan,
 0014, Republic of Armenia}
\author{Aleksander V. Bogdanov}
\affiliation{St. Petersburg State University, 7/9 Universitetskaya nab., St. Petersburg, 199034 Russia}

  \date{\today}
\begin{abstract}
Soon after the postulation of quarks by  Gell-Mann, Zweig and  Fritzsch the experimental confirmation of these sub-nucleon formations, Feynman, Ravndal and Kislinger proposed 
a relativistic three-quark model of the nucleon to study its internal structure and state. Despite the obvious progress in describing the internal motion of a system with confinement 
of quarks in a nucleon, it should be noted that the model turned out to be insufficiently realistic for a number of reasons. In particular, the model does not take into account the 
following cornerstone properties of QCD, namely:
{\bf a})  asymptotic freedom on small scales, {\bf b}) spontaneous breaking of chiral symmetry,
{\bf{c}}) the formation of a nucleon with a mass of about 1 GeV from the masses of up and down quarks of several MeV and
{\bf{d}}) color confinement of the nucleon. The disadvantages of this model are associated with the failure to describe the processes of gluon exchange between quarks, the 
influence of continuously formed pairs of quarks and antiquarks (quark sea) on valence quarks, as well as the self-interaction of colored gluons.

To eliminate the above shortcomings of the model, the problem of self-organization of a three-quark dynamical system immersed in a colored quark-antiquark sea is considered within the 
framework of complex probabilistic processes that satisfy the stochastic differential equation of  Langevin-Kline-Gordon-Fock type. Taking into account the hidden symmetry of 
the internal motion of a dynamical system, a mathematically closed non-perturbative approach has been developed, which makes it possible to construct the mathematical expectation 
of the wave function and other parameters of the nucleon in the form of multiple integral representations. The developed approach can be especially useful for studying the state
of nucleons in critical states, which occurs, for example, in massive and dense stellar formations such as neutron stars, etc.

\end{abstract}

\maketitle
\section{I\lowercase{ntroduction}}
\label{01}
In the observable universe, ordinary matter, and star formations in particular, are composed primarily of strongly interacting particles of protons and neutrons called nucleons. In this regard, 
one of the most important tasks of modern nuclear physics is the study of the structure of the nucleon and its excited states from the point of view of effective degrees of freedom and, at a 
more fundamental level, the emergence of these  from QCD \cite{Isgur}. After postulating the quark structure of strongly interacting particles by Gell-Man, Zweig and  Fritzsch \cite{Gel,Zweig,Fritzsch},  
based on the ideas of symmetry and invariance in the system of particles and fields, Feynman, Ravndahl and Kislinger \cite{Fey} was proposed a three-quark representation of nucleons within 
the framework of a four-dimensional relativistic quantum oscillator model. Note that the main idea underlying the three-quark nucleon model is that quarks, interacting through the potential 
of a four-dimensional harmonic oscillator, cannot move away from each other and become free.  In a series of papers \cite{Lipes,Kim,Kim1}, the authors discussed a quantum harmonic 
oscillator formalism to study the important features of hadronic structures in relativistic quark models. An important achievement of these models is that they allow a covariant-probabilistic 
interpretation of the wave functions under consideration  \cite{Kim2}. 

Despite the obvious successes of the relativistic model of a four-dimensional oscillator in describing the structure and internal motion of a nucleon with the effect of quark confinement, it still 
remains insufficiently realistic. The fact is that the nucleon model under consideration does not take into account the continuous processes of colored gluon exchange between quarks. Moreover,
 the model does not take into account the spontaneous breaking of chiral symmetry, which is responsible for the generation of nucleon mass from more elementary light quarks. The latter is 
 obviously a strong simplification of the problem.  The difficulties of the model become even more obvious when we have to consider nucleons in nuclei or dense and superdense stellar 
 formations. Recall that in this case the main characteristics of the processes of gluon exchange between quarks, as well as the properties of the quark sea, change, which in turn directly 
 affects the structure and other parameters of nucleons.

 To solve a number of the above problems, this paper considers a relativistic three-quark dynamical system immersed in colored gluon fields, which, in turn, generate a quark-antiquark sea.  
At the same time, we describe the interactions between quarks using a four-dimensional harmonic oscillator, which ensures confinement of quarks, their asymptotic freedom at short distances 
and chiral symmetry. We formulate the mathematical problem in terms of a complex probabilistic process that satisfies a  \emph{stochastic differential Equation} (SDE) of the Langevin-Klein-Gordon-Fock 
type. Note that this formulation of the problem allows us to take into account both elastic and inelastic processes of gluon exchange between quarks as well as self-action,  type of gluon-antigluon  
interactions. It is shown that for the case when fluctuations of the color quark-antiquark sea are characterized by complex processes of the Markov-Gauss type, it is possible to construct a mathematically 
closed non-perturbative theory of the nucleon with additional six-dimensional compact sub-spaces.  In particular, when calculating the mathematical expectation of various parameters of the nucleon, 
we perform averaging over additional sub-spaces, which breaks the chirial symmetry \cite{Gerald} but at the same time maintains the color of the nucleon. Recall that for all parameters of the nucleon we obtain 
double integral representations, where the integrand contains a function that is a solution to a system of two coupled second-order \emph{partial differential equations} (PDEs).

In conclusion, we note that considering the nucleon as a complex self-consistent system of ``three quarks + a sea of quarks-antiquarks'', as a problem of self-organization within the framework of 
the developed concept, allows us to go beyond the framework of perturbation theory, which is very important for obtaining new non-trivial results in the field of \emph{quantum chromodynamics} (QCD),
which is essentially a non-perturbative theory.

 The manuscript is organized as follows:\\
In Section 2,  briefly outlines the well-known formulation of the problem of the internal motion of a nucleon as a three-quark dynamical system within the framework of the Klein-Gordon-Fock equation 
using a four-dimensional model of a relativistic oscillator for quark interactions   \cite{Lipes}. An exact relativistically invariant solution of the wave function of the internal motion of a nucleon is presented.

In Section 3, the problem of the internal motion of a nucleon immersed in a colored quarks sea is mathematically formulated in the framework of a complex probabilistic process that satisfies the 
Langevin-Klein-Gordon-Fock type SDE. It is shown that a complex probabilistic process in the model of a four-dimensional relativistic oscillator, after a convenient transformation of coordinates, is 
written in factorized form, as a product consisting of three independent functions.

In Section 4,  a system of stochastic Equations for gluon fields is derived taking into account the synchronization of four-dimensional events in the Minkowski space. The conditions of complex stochastic 
processes for generators of colored gluon fields, in the form of Markov-Gaussian processes, are determined.

In Section 5, taking into account the SDE system, an Equation for the distribution of gluon fields in the limit of statistical equilibrium is derived. It is shown that the solution for the field distribution is 
factorized into the product of three two-dimensional distributions, each of which describes the states of gluon fields of a certain color and anticolor.  It is shown that the additional six-dimensional 
sub-space generated by the SDEs system is factorized as a direct product of three two-dimensional sub-spaces.

In Section 6, the geometric and topological features of the emerging two-dimensional sub-spaces are analyzed. It is shown that each of the subspaces is generated by an algebraic equation of the 
fourth degree and is described by non-commutative geometry. It is also proven that the typological features of these manifolds are characterized by the Betti number $n\leq4$.

In Section 7, using a two-dimensional distribution Equation for gluon fields, a Fokker-Planck measure of the functional space is constructed.

In Section 8, the mathematical expectation of the total wave function of a three-quark dynamical system immersed in a colored quark-antiquark sea is defined in the form of a functional integral 
representation. Using the generalized Feynman-Kac theorem, functional integrals were calculated. As a result, a factorized representation in the form of three double integrals 
was obtained for the mathematical expectation of the wave function of a nucleon internal motion.

In Section 9, using the mathematical expectation of the total wave function of a nucleon, its radius and mass are determined depending on the constants characterizing the fluctuation powers of 
colored gluon fields. Two independent equations are derived that make it possible to uniquely calculate two independent constants characterizing the gluon fields of a nucleon when the nucleon is in a free state.

In Section 10, discusses in detail the possibilities of representation for a more correct description of the quantum state of the nucleon, taking into account the processes of gluon exchange between 
quarks. The issues of the state of a nucleon in the case of their immersion in dense and superdense stellar formations are discussed in the light of changes in the spectrum and power of fluctuations
of gluon fields.

In Section 11, discusses the issues of color synchronization of valence quarks or the problem of preserving the white color of the nucleon, which, in turn, is closely related to the problem of the 
three-particle interaction of valence quarks.

 
\section{R\lowercase{elativistic three-quark  dynamical system}}
As is known, the main assumption about the dynamics of the internal motion of a nucleon is that the motion of three quarks is described by the relativistic Klein–Gordon–Fock equation in the light 
front formalism, where the interaction between quarks is carried out through a four-dimensional harmonic potential ensuring confinement of quarks in the nucleon \cite{Fey}. Taking this work into 
account, the following Lorentz-invariant  Equation was proposed to describe the state of a three-quark dynamical system \cite{Lipes}:
 \begin{equation}
\Bigl\{\sum_{\zeta=x_a,x_b,x_c}\Box_{\zeta} -\frac{1}{27}\,{\Omega_0^2} \bigl[({{x}}_a-{{x}}_b)^2+({{x}}_a-{{x}}_c)^2+
({{x}}_b-{{x}}_c)^2\bigr]+m_0^2\Bigr\}\Psi^{(0)} (x_a,x_b,x_c)=0,
\label{q1.01}
\end{equation}
where ${{x}}_a,{{x}}_b$ and ${{x}}_c$ are the four-dimensional space-time coordinates of quarks $\bf{a},\bf{b}$ and $\bf{c}$ (see Fig. 1), in addition, $\Omega_0$ is the some constant,
$m_0$ denotes the sum of the rest masses of three quarks,  and $\Box_\zeta$ denotes the d'Alembert operator acting on the $\zeta$-\emph{th} quark: 
$$\Box_\zeta=\partial^2_t-\nabla^2(x_\zeta,y_\zeta,z_\zeta)=\partial^2_t-\partial^2_{x_\zeta}-\partial^2_{y_\zeta}-\partial^2_{z_\zeta}.$$
Recall that below all calculations will be carried out in units; $\hslash=c=1.$

By performing the following coordinate transformations:
\begin{equation}
u=\frac{1}{\sqrt{3}}({x}_a+{x}_b+{x}_c),\quad {v}=\frac{1}{\sqrt{2}}({x}_a-{ x}_b),\quad { w}=\frac{1}{\sqrt{6}}({x}_a+{x}_b-2{x}_c),
\label{q1.t02}
\end{equation} 
the Equation (\ref{q1.t02}) can be reduced to diagonal form:
\begin{equation}
 \Bigl\{\sum_{\xi={u},{ v},{ w}}\Box_{\xi}+m_0^2-\frac{1}{9}\Omega_0^2({v}^2+{w}^2)\Bigr\}\Psi ({u},{v},{w})=0,
\label{q1.02}
\end{equation}
where $\xi=(\xi_0,\xi_1,\xi_2,\xi_3)$ and, accordingly, the d'Alembert operator in new coordinates has the form $\Box_{\xi}=\partial_{\xi_0}^2-\sum_{k=1}^3\partial ^2_{ \xi_k}.$
\begin{figure} 
\centering
\includegraphics[width=100mm]{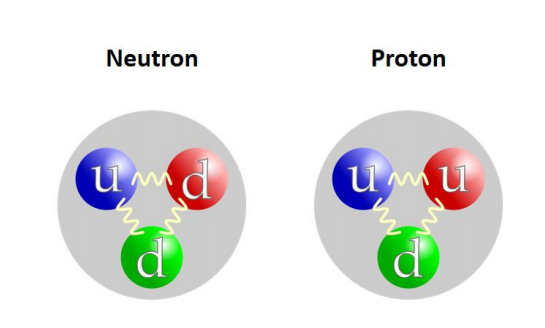}
\caption{\emph{The pictures show two nucleons (proton and neutron) in the form of valence quarks, located in different quantum states, i.e. in different colors. Conventionally, we will assume that the 
$\bf a$-quark is depicted in blue, the $\bf b$-quark in red, and the $\bf c$-quark in green.}
\label{overflow12}}
\end{figure}

Now representing the solution of  the  Equation (\ref{q1.02}) in factored form:
\begin{equation}
\Psi({u},{v},{w})=\Psi_1(u)\Psi_2(v)\Psi_3(w),
\label{q1.z02}
\end{equation}
from  (\ref{q1.02})  we obtain three new Equations:
\begin{eqnarray}
\bigl[\Box_{u} +\lambda_{01}\bigr]\Psi_1(u)=0, \qquad\qquad\qquad
\nonumber\\
\bigl[\Box_{v} +\lambda_{02}\,-(1/9)\Omega_0^2\,v^2\bigr]\Psi_2(v)=0, 
\nonumber\\
\bigl[\Box_{w} +\lambda_{03}-(1/9)\Omega_0^2w^2\bigr]\Psi_3(w)=0, 
\label{q1.k01}
\end{eqnarray}
where $m_0^2=\lambda_{01}+\lambda_{02}+\lambda_{03}$.

The solution to the first Equation of the system (\ref{q1.k01}) can be represented as:
\begin{equation}
\Psi_1(u)=\exp\bigl\{-{i}u\cdot P/{\sqrt{3}}\bigr\},
\label{q1.K0wt3}
\end{equation}
where $P^\nu (\nu=0,1,2,3)$ is a four vector. \\
Substituting this solution into the first Equation of the system (\ref{q1.k01}), it is easy to find that $P^2=3\lambda_{01}.$ In addition, from (\ref{q1.k01}) follows that $\lambda_{01}=\lambda_{02}=(1/3)\Omega^2_0=(1/3)m_0^2.$ 

The remaining two equations describe the quantum motions of two independent four-dimensional oscillators, described by the wave functions $\Psi_2(v)$ and $\Psi_2(w)$, respectively. In the rest  frame, where 
$P^\nu=(m_0,{\bf0})=(m_0,0,0,0)$, the \emph{ground state} wave function normalized in relative coordinates $v$ and $w$, can be written as:
\begin{equation}
 \Psi({u},{v},{w};{\bf 0})=\biggl(\frac{\Omega_0}{3\pi}\biggr)^2\exp\biggl\{-{i}\frac{m_0u_0}{{\sqrt{3}}}-\frac{\Omega_0}{6}\bigl(v_0^2+{\bf v}^2+w_0^2+{\bf w}^2\bigr)\biggr\},
\label{q1.k0t3}
\end{equation}
which may be written in covariant form:
$$
 \Psi({u},{v},{w};{\bf P})=\biggl(\frac{\Omega_0}{3\pi}\biggr)^2\exp\biggl\{-{i}\frac{m_0u_0}{{\sqrt{3}}}-\frac{\Omega_0}{6m^2_0}\Bigl(2\bigl[{(P\cdot v)^2}+{(P\cdot w)^2}\bigr]-m^2_0\bigl[{v}^2 +{w}^2\bigr]\Bigr)\biggr\}.
$$
As for the wave function of the excited state, as shown in the work \cite{Lipes}, it can be represented in the form:
\begin{equation}
 \Psi_{{\bf n}\,{\bf m}}(u,v,w)=N_{{\bf n}} N_{{\bf m}}\biggl[\,\prod_{k=1,2,3}H_{n_k}(v_k)H_{m_k}(w_k)\biggr]  \Psi({u},{v},{w};{\bf0}),  \quad N_{{\bf n}}=N_{n_1}N_{n_2}N_{n_3}, 
\label{q1.k0zt3}
\end{equation}
where $N_{n_k}$  is the normalization constant of  the one-dimensional oscillator, ${\bf n}=(n_1, n_2, n_3)$ the set of quantum number, and $H_n(x)$ denotes the Hermite polynomials.
Recall that the wave function of a four-dimensional oscillator has the form:
\begin{equation}
\Psi_2({\bf n};v)=N_{\bf n}\prod_{k=1,2,3}H_{n_k}(v_k)\exp\Bigl\{-\frac{\Omega_0}{6}v_k^2\Bigr\}.
\label{q1.k0zt3}
\end{equation}
Note that we obtain exactly the same function for another oscillator described by the function $\Psi_3({\bf m};w)$.

As can be seen from expressions (\ref{q1.k0t3}) and (\ref{q1.k0zt3}), the wave function of the nucleon is localized along the coordinates $v$ and $w$, while the three-quark dynamical system performs
 translational motion along the coordinates $u$.
 
\section{T\lowercase{hree-quark dynamical system immersed in a colored quark-antiquark sea}}

Since nucleons consist of combinations of two types of light quarks $\bf u$ and $\bf d$ (see Fig. 1), it is natural to expect that the interaction of these quarks should be carried out by gluons of different colors.  
In particular, when the colors of two quarks are known, the interaction between them will be carried out by gluons of these colors and corresponding antigluons.  Note that by antigluon we mean the same 
gluon that has an anticolor. Having gone through all the combinations, it becomes obvious that a nucleon, consisting of three quarks of different colors without taking into account the spins of the quarks, is 
immersed in a six-color quark-antiquark sea  (see Fig. 2). Taking this into account, we must construct a consistent non-perturbative theory of self-organization of a quark system and its random multi-color 
quark-antiquark environment.  Based on the experience of studying similar problems of nonrelativistic quantum mechanics \cite{Ash}, we can rewrite Equation (\ref{q1.01}) in the following form:
\begin{eqnarray}
\Bigl\{\sum_{\zeta=x_a,x_b,x_c}\Box_{\zeta}  -\frac{1}{27}\,{\Omega_0^2} \bigl[({{x}}_a-{{x}}_b)^2+({{x}}_a-{{x}}_c)^2+
({{x}}_b-{{x}}_c)^2\bigr] +m^2(x_a,x_b,x_c)\Bigr\}\Psi=0,
\label{q2.01}
\end{eqnarray}
where $m^2(q_a,q_b,q_c)$ is some space-time complex random function, the form of which will be indicated below. Recall that the valence quarks $\bf a$, $\bf b$ and $\bf c$ (see Fig. 1) in a nucleon are 
in three different quantum states or colors, so the mathematical expectation of the mixed color of a nucleon should  be white.
\begin{figure} 
\centering
\includegraphics[width=70mm]{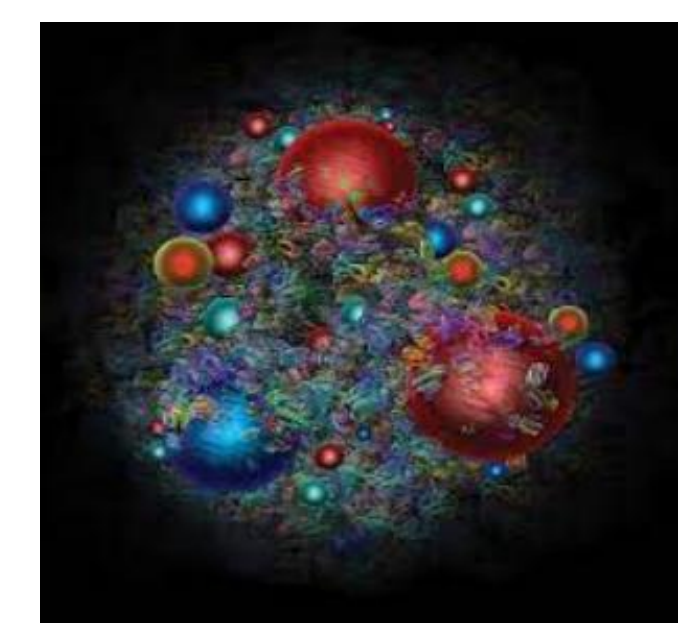}
\caption{\emph{Three valence quarks are immersed in a colored quark-antiquark sea inside a single proton. As the numerical simulation of the problem (see for example \cite{Gees} and \cite{Gaz}), as well 
as experimental studies,  show, with a decrease  in space-time scales, the powers and frequency of gluon fields fluctuations increase. As can be seen from the figure, for very short times the color symmetry 
of valence quarks can be violated.}
\label{overflow12}}
\end{figure}

Using coordinate transformations (\ref{q1.t02}) the Equation  (\ref{q2.01})  can be written as:
\begin{equation}
\Bigl\{\sum_{\xi=u,v,w}\Box_{\xi} -\frac{1}{9}{\Omega_0^2}\bigl( v^2+w^2\bigr)+\breve{m}^2(u,v,w)\Bigr\}\breve{\Psi}(u,v,w)=0,
\label{q2.02}
\end{equation}
where $\breve{m}^2(u,v,w)=m^2(x_a,x_b,x_c).$

To clarify the function $\breve{m}^2(u,v,w)$, we will assume that it can be represented in the form:
\begin{equation}
 \breve{m}^2(u,v,w)=m_0^2+\{{\lambda}\}, \qquad \{{\lambda}\}=\lambda_1(s_u)+\lambda_2(s_v)+\lambda_3(s_w),
\label{q2.z02}
\end{equation}
where $\lambda_1(s_u)$, $\lambda_2(s_v)$ and $\lambda_3(s_w)$ some complex random processes depending on an proper space-time events:
\begin{equation}
 s_\xi=\bigl({\xi_0^2-\sum_{\delta=x,y,z}\xi_\delta^2}\bigr)^{1/2},\qquad \xi=u,v,w.
\label{q2.zt02}
\end{equation}
Recall that since we are considering a relativistic problem, it is natural to use a 4-vector, which defines a chronologized sequence of space-time events in Minkowski space, as a parameter describing the 
evolution of a dynamical system.

Finally, based on the symmetry of Equation (\ref{q2.02})-(\ref{q2.z02}), the solution to the total wave function of the system can be represented as the following product:
\begin{equation}
\breve{\Psi}(u,v,w|\{{\lambda}\})=\breve{\Psi}_1\bigl(u,\lambda_1(s_u)\bigr) \cdot\breve{\Psi}_2\bigl(v,\lambda_2(s_v)\bigr)\cdot\breve{\Psi}_3\bigl(w,\lambda_3(s_w)\bigr).
\label{q2.03}
\end{equation}
Substituting (\ref{q2.03}) into (\ref{q2.02}), taking into account (\ref{q2.z02}), we obtain three new equations:
\begin{eqnarray}
\bigl[\Box_{u} +\lambda_{01}+\lambda_1(s_u)\bigr]\breve{\Psi}_1\bigl(u,\lambda_1(s_u)\bigr)=0, \qquad\qquad\qquad\,
\nonumber\\
\bigl[\Box_{v}\, +\lambda_{02}\,-({1}/{9})\Omega_0^2\,v^2+\lambda_2(s_v)\bigr]\breve{\Psi}_2\bigl(v,\lambda_2(s_v)\bigr)\,=0, 
\nonumber\\
\bigl[\Box_{w} +\lambda_{03}-(1/9)\Omega_0^2w^2+\lambda_3(s_w)\bigr]\breve{\Psi}_3\bigl(w,\lambda_3(s_w)\bigr)=0.
\label{q2.04}
\end{eqnarray}
It should be noted that the system of equations (\ref{q2.04}), due to the presence of random generators $\lambda_k(s_\xi), (k=1,2,3)$ in them in this form, still requires definition and reduction to canonical form, 
i.e. to the form of a first-order differential equation of Langevin type.

Now our main task will be to construct the mathematical expectation of the total wave function of a nucleon on the light cone, taking into account gluon exchanges between quarks and the influence of the quark-antiquark sea:
\begin{equation}
\overline{\Psi}(x_a,x_b,x_c)=\overline{\breve{\Psi}}(u,v,w)=\mathbb{E}[\breve{\Psi}(u,v,w|\{\lambda\})],
\label{q2.W03}
\end{equation}
where $\mathbb{E}[\cdot\cdot\cdot]$ denotes the mathematical expectation of a random variable, in addition, recall that the sets $(x_a,x_b,x_c)$ and $(u,v,w)$ denote 4 -vectors of Minkowski space-time.

\section{E\lowercase{quations of motion of gluon fields under the influence of valence quarks}}
In view of the foregoing, the solution of each of the Equations (\ref{q2.04}) can be represented as:
\begin{eqnarray}
\breve{\Psi}_k(\xi,s_\xi|\lambda_k(s_\xi))=\Psi_k(\xi)\exp\biggl(\int_0^{s_\xi}\Lambda_k(\xi,s')ds'\biggr),\qquad k=1,2,3,  
\label{q3.01}
\end{eqnarray}
where $\Psi_k(\xi)$ is a regular function, a solution to one of the Equations of the system (\ref{q1.k01}) and $\Lambda_k(\xi,s_\xi)$ denotes a complex probabilistic process, 
which can be conveniently represented as a sum consisting of real and imaginary terms:
\begin{equation}
\Lambda_1 =\sum_{j=1,2}i^{j-1}\phi_j(s_u|u),\quad  \Lambda_2 =\sum_{j=1,2}i^{j-1}\varphi_j(s_v|v),
 \quad
 \Lambda_3 =\sum_{j=1,2}i^{j-1}\theta_j(s_w|w).
\label{q3.0x3}
\end{equation}

Substituting a solution of the form (\ref{q3.01}) into the corresponding Equations of the system (\ref{q2.04}), we obtain:
\begin{equation}
\begin{cases}
\dot{\Lambda}_1 +a_1\Lambda_1+b_1\Lambda^2_1+c_1\lambda_1(s_u)=0, 
\\
\dot{\Lambda}_2 +a_2\Lambda_2+b_2\Lambda^2_2+c_2\lambda_2(s_v)=0, 
\\
\dot{\Lambda}_3 +a_3\Lambda_3+b_3\Lambda^2_3+c_3\lambda_3(s_w)=0,
\label{q3.02}
\end{cases}
\end{equation} 
where
\begin{equation}
\dot{\Lambda}_k=\partial\Lambda_k /\partial s_\xi,   \qquad a_k=\frac{2\xi\cdot \nabla_M  \ln\Psi_k-2-s_\xi}{\xi_0-\sum_{k=1}^3\xi_k}, \qquad b_k=c_k=\frac{s_\xi}{\xi_0-\sum^3_{k=1}\xi_k}.
\label{q3.ZX3}
\end{equation}
Note that the operator $\nabla_M=(\partial_{\xi_0},-\partial_{\xi_1},-\partial_{\xi_2},-\partial_{\xi_3})$ denotes the gradient in Minkowski  space-time $\mathbb{R}^4$.

Now we define random functions $\lambda_k(s_\xi)$ that characterize the properties of colored and anticolored gluon fields, representing them as the sum of real and imaginary terms:
\begin{equation}
\lambda_k(s_\xi)=f^{(r)}_k(s_\xi)+if_k^{(i)}(s_\xi),\qquad k=1,2,3.
\label{q3.0z3}
\end{equation}
For definiteness, we will assume that these  functions satisfy Markov-Gaussian random processes or \emph{white noise} correlation relations:
\begin{eqnarray}
\mathbb{E}\bigl[f_j^{(\upsilon)}(s_\xi)\bigr] =0,\qquad \mathbb{E}\bigl[f_j^{(\upsilon)}(s_\xi)f_j^{(\upsilon)}(s'_\xi)\bigr]
= 2\varepsilon_j^{(\upsilon)}\,\delta (s_\xi-s'_\xi),\qquad j=1,2,
\label{q3.04}
\end{eqnarray}
where $ \upsilon=(i,r)$. 

We assume that the random generators $f_j^{(r)}$ and $f_j^{(i)}$ characterize elastic and inelastic processes of exchange of gluons and antigluons of a given color between two specific quarks.

Finally, using the expressions (\ref{q3.01})-(\ref{q3.0z3}), we can obtain the following system of six non-linear Langevin-type SDEs:
\begin{equation} 
\begin{cases}
\,\dot{\phi}_1 +a^{(r)}_1\phi_1-a^{(i)}_1\phi_2+b_1(\phi^2_1-\phi^2_2)+c_1f^{(r)}_1(s)=0, 
\\
\qquad\dot{\phi}_2 +a^{(i)}_1\phi_1+a^{(r)}_1\phi_2+2b_1\phi_1\phi_2+c_1f^{(i)}_1(s)=0, 
\\
\,\dot{\varphi}_1 +a^{(r)}_2\varphi_1-a^{(i)}_2\varphi_2+b_2(\varphi^2_1-\phi^2_2)+c_2f^{(r)}_2(s)=0, 
\\
\qquad\dot{\varphi}_2 +a^{(i)}_2\phi_1+a^{(r)}_2\varphi_2+2b_2\varphi_1\varphi_2+c_2f^{(i)}_2(s)=0, 
\\
\,\,\,\dot{\theta}_1\, +a^{(r)}_3 \theta_1\,-a^{(i)}_3\theta_2\,+b_3(\theta^2_1-\theta^2_2)\,+c_3f^{(r)}_3(s)=0, 
\\
\qquad\,\dot{\theta}_2\, +a^{(i)}_3\theta_1\,+\,a^{(r)}_3\theta_2+\,2b_3\theta_1\theta_2+\,c_3f^{(i)}_3(s)=0. 
\label{q3.0zt2}
\end{cases}
\end{equation}
Recall that for further study of the system of Equations (\ref{q3.0zt2})  we synchronized the evolutionary parameters, that is, we took  the smallest parameter among them $s=\min\{s_u,s_v,s_w\}$.

Thus, the system of Equations (\ref{q3.0zt2}) describes colored gluon fields under the influence of three valence quarks, and also takes into account the self-actions of gluons. Now it is important to use the 
SDEs  (\ref{q3.0zt2})  to obtain a regular equation that describes the distribution of gluon fields in the limit of statistical equilibrium. Note that this is fundamentally important for a consistent analytical 
construction of the problem.

\section{D\lowercase{istribution of colored gluon fields in the limit of statistical equilibrium }}
When gluons are exchanged between two valence quarks of different colors, since the overall white color of the nucleon must be conserved, this must affect the dynamics of the third valence quark to 
preserve the color of the nucleon. It follows that any processes of gluon exchange between two quarks make the interaction in the nucleon three-particle. The latter means that the probabilities of the 
distribution of gluon fields of all six colors should be interconnected and combined. Taking into account the above, it is necessary to represent the distribution of gluon fields in the following form:
\begin{equation}
\mathcal{P}({\bm \vartheta},s|{\bm\vartheta}_{0},s_0)=\Bigr\langle\prod_{j=1,2\,\,}\prod_{{\vartheta}=\phi,\varphi,\theta}\delta\bigl(\vartheta_j(s)-\vartheta_{0j}\bigr)\Bigr\rangle,
\label{q4.01}
\end{equation}
where  ${\bm\vartheta}(s)=\bigl\{\bm\phi(s),\bm\varphi(s),\bm\theta(s)\bigr\}\in {\Xi}_{\{{\bm\vartheta}(s)\}}$  and  $\langle\cdot\cdot\cdot\rangle$ denotes the functional integration over the functional space
 ${\Xi}_{\{{\bm\vartheta}(s)\}}$, whose measure will be defined below, in addition, $\vartheta_{0j}=\vartheta_j(s_0)$  denotes the field value in $s_0=0$.

Using the SDE system (\ref{q3.04}), it is possible to strictly prove that the probability distribution of gluon fields satisfies the following Fokker–Planck type PDE (see \cite{Kly,Ash1}):
\begin{equation}
\frac{\partial\mathcal{P}}{\partial s}=\widehat{\mathcal{L}}({\bm \vartheta},s|u,v,w)\mathcal{P},
\label{q4.01}
\end{equation}
where the evolution operator $\widehat{\mathcal{L}}({\bm \vartheta},s|u,v,w)$ has the following form:
\begin{eqnarray}
\widehat{\mathcal{L}} = \biggl\{\biggl(\bar{\varepsilon}_1^{(r)}\frac{\partial^2}{\partial \phi_1^2}+\bar{\varepsilon}_1^{(i)}\frac{\partial^2}{\partial \phi_2^2}\biggr)+
 \biggl(\bar{\varepsilon}_2^{(r)}\frac{\partial^2}{\partial \varphi_1^2}+\bar{\varepsilon}_2^{(i)}\frac{\partial^2}{\partial \varphi_2^2}\biggr)+ \biggl(\bar{\varepsilon}_3^{(r)}\frac{\partial^2}{\partial \theta_1^2}+
 \bar{\varepsilon}_3^{(i)}\frac{\partial^2}{\partial \theta_2^2}\biggr)\biggr\}\,
 \nonumber\\
+\sum^2_{j=1}\biggl\{\frac{\partial}{\partial \phi_j} \sigma_j({\bm\phi},s |u)+ \frac{\partial}{\partial \varphi_j} \pi_j({\bm\varphi},s |v)+ \frac{\partial}{\partial \theta_j} \omega_j({\bm\theta},s |w)\biggr\}.
\label{q4.02}
\end{eqnarray}
where $\bar{\varepsilon}_k^{(\upsilon)}=c_k^2\varepsilon_k^{(\upsilon)},\, (k=1,2,3)$.
 
Note that in the Equation (\ref{q4.02})  the follwing notations are made $ {\bm\phi}=(\phi_1,\phi_2),$ ${\bm\varphi}=(\varphi_1,\varphi_2)$ and ${\bm \theta}=(\theta_1,\theta_2)$, in addition:
\begin{eqnarray}
 \sigma_1({\bm\phi},s|u)=a_1^{(r)}\phi_1-\,a_1^{(i)}\phi_2+b_1[\phi^2_1-\phi_2^2],\quad\sigma_2({\bm\phi},s|u)=a_1^{(i)}\phi_1+a_1^{(r)}\phi_2+2b_1\phi_1\phi_2, 
 \nonumber\\
\pi_1({\bm\varphi},s|v)=a_2^{(r)}\varphi_1-a_2^{(i)}\varphi_2+b_2[\varphi^2_1-\varphi_2^2],\quad \pi_2({\bm\varphi},s|v)=a_2^{(i)}\varphi_1+a_2^{(r)}\varphi_2+2b_2\varphi_1\varphi_2, 
 \nonumber\\
 \omega_1({\bm\theta},s|w)=a_3^{(r)}\,\theta_1-a_3^{(i)}\theta_2+\,b_3[\theta^2_1\,-\,\theta_2^2],\quad  \omega_2({\bm\theta},s|w)=a_3^{(i)}\theta_1\,+\,a_3^{(r)}\theta_2+\,2b_3\,\theta_1\theta_2,
\label{q4.0z2}
\end{eqnarray}
where $a^{(r)}_k=Re\{a\}$ and $a^{(i)}_k=Im\{a\}, \, (k=1,2,3).$

It is important to note that the Equation (\ref{q4.01})-(\ref{q4.02})  takes into account gluon-antigluon interactions.  Recall that this is reflected in the mixed terms $\sigma_2({\bm\phi},s|u),\, \pi_2({\bm\varphi},s|v)$
and $\omega_2({\bm\theta},s|w)$. It is important to note that since colored gluon fields generate a sea of colored quarks-antiquarks, the indicated distributions also 
describe a quark sea.

The symmetry of the Equation (\ref{q4.01})-(\ref{q4.02}) allows us to represent its solution in factorized form:
\begin{equation}
\mathcal{P}({\bm\vartheta},s|u,v,w)=\mathcal{P}_{{\bm\phi}}({\bm\phi},s|u)\cdot\mathcal{P}_{\bm\varphi}({\bm\varphi},s|v)\cdot\mathcal{P}_{\bm\theta}({\bm\theta},s|w),
\label{q4.03z}
\end{equation}
where each of the probability distributions $\mathcal{P}_{\bm\phi}({\bm\phi},s|u),\,\,\mathcal{P}_{\bm\varphi}({\bm\varphi},s|v)$ and  $\mathcal{P}_{{\bm\theta}}({\bm\theta},s|w)$ is defined on the corresponding
two-dimenssional manifold. In other words, the additional six-dimenssional sub-space $\Xi^6_{\{{\bm\vartheta}\}}$ in the limit of the statistical equilibrium can be represented as the following decomposition:  
\begin{equation}
\Xi^6_{\{{\bm\vartheta}\}}\cong\Xi^2_{\{{\bm\phi}\}}\bigotimes\Xi^2_{\{{\bm\varphi}\}}\bigotimes\Xi^2_{\{{\bm\theta}\}}.
\label{q4.w03z1}
\end{equation}
If the nucleons are free, then we assume that the powers of gluon fluctuations are relatively small, i.e. $\bar{\varepsilon}_k^{(r)},\bar{\varepsilon}_k^{(i)}\sim1$ for all $k=1,2,3$. In this case, we can safely assume
that all additional two-dimensional sub-spaces are Euclidean;
$
\Xi^2_{{\bm\phi}}\cong\Xi^2_{{\bm\varphi}}\cong\Xi^2_{{\bm\theta}} \cong \mathbb{R}^2 = (-\infty,+\infty){\bf\times}(-\infty,+\infty).
\label{q4.w03z}
$

An important feature of the equation (\ref{q4.02}), which describes the distribution of fields of colored gluons, or rather the sea of quarks-antiquarks, in the limit of statistical equilibrium, is quantized.
Recall that this follows from the dependence of the function $a_k(\xi,s_\xi), \,(k=1,2,3)$ on the quantum state of quarks (see the definition of the function (\ref {q3.02})), which is characterized by six quantum numbers. 

\section{G\lowercase{eometric and topological features of the additional two-dimensional sub-spaces}}
In the case when the  fluctuation's powers for some reason take on large values; $\bar{\varepsilon}_k^{(r)},\,\bar{\varepsilon}_k^{(i)}\gg1$ for all $k=1,2,3$ it is necessary to conduct a comprehensive analysis to 
identify the geometric and topological features of the two-dimensional manifolds; $\Xi^2_{\{{\bm\chi}\}},\,\,({\bm\chi}={\bm\phi},{\bm\varphi},{\bm\theta})$.

{\bf Theorem}. \emph{If a dynamical system is described by any of the SDEs from (\ref{q3.02}), then in the limit of statistical equilibrium the functional space ${\Xi}_{\{{\bm \chi}(s )\}},\, \{{\bm\chi}(s)=
{\bm\phi}(s), {\bm\varphi}(s),{\bm\theta}(s)\} $ is compactified into the two-dimensional manifold  $\Xi^2_{\{{\bm\chi}\}},\,\,({\bm\chi}={\bm\phi},{\bm\varphi},{\bm\theta})$, which is characterized by 
non-commutative geometry.}

\begin{proof}

We will consider the distribution of two-component gluon fields or color-anticolor fields. In particular, the fields probability density Equation ${\bm\phi}=(\phi_1,\phi_2)$ can be represented as (see {\bf{Appendix}}):
\begin{equation}
\frac{\partial\mathcal{P}_{\bm\phi}}{\partial s} =\widehat{\mathcal{L}}^{(0)}_{\bm\phi}\mathcal{P}_{\bm\phi}, \qquad \widehat{\mathcal{L}}^{(0)}_{\bm\phi}=\biggl\{\biggl(\bar{\varepsilon}_1^{(r)}
\frac{\partial^2}{\partial \phi_1^2}+\bar{\varepsilon}_1^{(i)}\frac{\partial^2}{\partial \phi_2^2}\biggr)+\sum^2_{j=1}\frac{\partial}{\partial \phi_j} \sigma_j({\bm\phi},s |u)\biggr\}.
\label{qt4.02w}
\end{equation}
Let us write the same equation (\ref{qt4.02w}) in tensor form for further analysis~\cite{Ash1,Jost}):
\begin{equation}
\frac{\partial\mathcal{P}_{\bm\phi}}{\partial s}=\Bigl\{ \nabla^2+k_{\bm\phi}(\phi_1,\phi_2,s)\Bigr\}\mathcal{P}_{\bm\phi},\qquad \nabla^2=\frac{1}{\sqrt{|g|}}
\sum^2_{i,j=1}\frac{\partial}{\partial \phi^i}\biggl(\sqrt{|g|}g^{ij}\frac{\partial}{\partial \phi^j}\biggr),
\label{nw3.01}
\end{equation}
where  $\phi_1=\phi^1$ and $\phi_2=\phi^2$, in addition, $k_{\bm\phi}=2a^{(r)}_1+4b_1\phi_1$.

To find the elements of the metric tensor, we write the two-dimensional Laplace-Beltrami operator $\nabla^2$ in explicit form:
\begin{eqnarray}
\nabla^2=g^{11}\frac{\partial^2}{\partial \phi_1^2} +
\frac{1}{\sqrt{|g|}}\biggl[\frac{\partial}{\partial \phi_1}\Bigl(\sqrt{|g|}g^{11}\Bigr)+\frac{\partial}{\partial \phi_2}
\Bigl(\sqrt{|g|}g^{21}\Bigr)\biggr]\frac{\partial}{\partial \phi_1}+g^{12}\frac{\partial^2}{\partial \phi_1\partial \phi_2}\,\,
\nonumber\\
\quad\quad+\,\,g^{22}\frac{\partial^2}{\partial \phi_2^2}
+\frac{1}{\sqrt{|g|}}\biggl[\frac{\partial}{\partial \phi_2}\Bigl(\sqrt{|g|}g^{22}\Bigr)+\frac{\partial}{\partial \phi_1}
\Bigl(\sqrt{|g|}g^{12}\Bigr)\biggr]\frac{\partial}{\partial \phi_2} +g^{21}\frac{\partial^2}{\partial \phi_2\partial \phi_1}.
\label{nw3.02}
\end{eqnarray}
Comparing the operator written in the forms (\ref{q4.02}) and (\ref{nw3.02}), and requiring the equality of the corresponding terms in the Equations, we find:
\begin{eqnarray}
g^{11}=\bar{\varepsilon}^{(r)} ,\quad g^{22}=\bar{\varepsilon}^{(i)},\quad g^{12}=-g^{21},
\quad
 g=g^{11}g^{22}-g^{12}g^{21}=\bar{\varepsilon}^{(r)}\bar{\varepsilon}^{(i)} +\bigl(g^{12}\bigr)^2.
 \label{q4.z04}
\end{eqnarray}
As can be seen from the expressions (\ref{q4.z04}), the metric tensor of the additional subspace $\Xi^2_{\{{\bm\phi}\}}$ is antisymmetric, which implies that the corresponding geometry is non-commutative.
A similar comparison of equations (\ref{nw3.02}) and (\ref{q4.02}) allows us to obtain the following first-order differential Equations for the non-diagonal element of the metric tensor $g^{12}= y$:
\begin{equation}
 \Biggl\{
\begin{array}{ll} 
\bar{\varepsilon}^{(r)}\eta\partial_{\phi_1} y- (1+y\eta)\partial_{\phi_2} y =\sigma_1({\bm\phi},s|{u}),
\\
\bar{\varepsilon}^{(i)}\eta\partial_{\phi_2}y+ (1+y\eta) \partial_{\phi_1} y=\sigma_2({\bm\phi},s|{u}),\qquad \partial_{x}=\partial/\partial x,
\end{array}
\label{q4.t04}
\end{equation}
where 
$$
\eta(y,s|u)=\frac{y}{\varepsilon^{(r)}\varepsilon^{(i)}+y^2}.
$$
Now our main task will be to use (\ref{q4.t04}) to obtain an algebraic Equation that allows us to determine the element of the metric tensor $y$.
\begin{figure} 
\centering
\includegraphics[width=70mm]{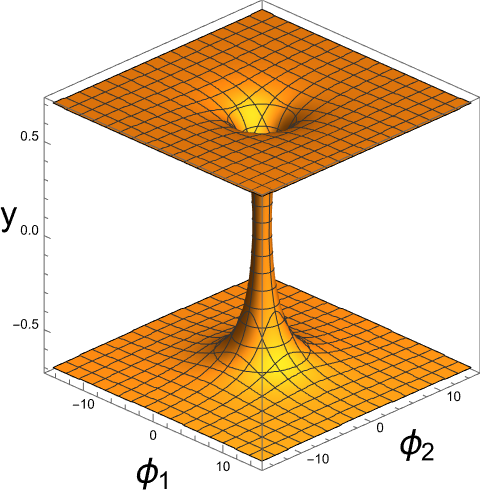}
\qquad
\includegraphics[width=70mm]{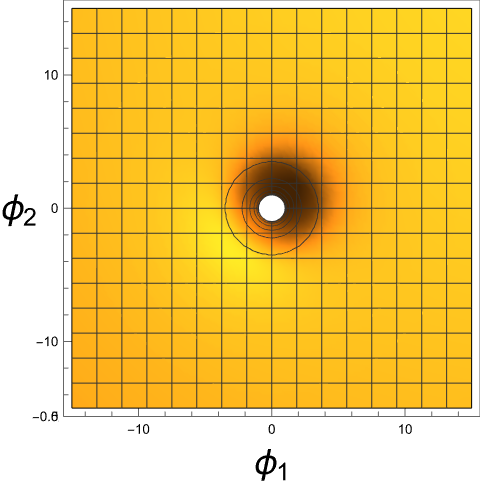}
\caption{\emph{The left figure is a three-dimensional plot of the asymmetric element of the metric tensor $g^{12}(\phi_1,\phi_2)$ in the $s\to\infty$ limit, calculated in the nucleon ground state, for 
the values of parameters $\varepsilon^{(r)} =\varepsilon^{(i)}=1$ and $a^{(r)}=0,\,\,a^{(i)}=b_1=1$. The right figure shows a two-dimensional projection of the submanifold $\Xi_{\{{\bm\phi}\}}^2$, 
where it is clearly seen that it has a singularity characterized by a Betti number of 1.}
\label{overflow12}}
\end{figure}

Using Equations (\ref{q4.t04}), we can find the following two expressions for the mixed second derivatives of the anti-symmetric element of the metric tensor:
\begin{eqnarray}
{y}_{12}=\frac{\partial^2\,{y}}{\partial{\phi}_1\partial{\phi}_2}\,=\,\frac{\bar{\epsilon}^{(i)}({\sigma}_{1;2}\eta\,+{\sigma}_{1}\eta_{2})+{\sigma}_{2;2}(1+{y}\eta)
+\sigma_2(y_2\eta+y\eta_{;2})}{a\eta^2+(1+{y}\eta)^2}
\nonumber\\
-2\frac{\bar{\epsilon}^{(i)}{\sigma}_1\eta+{\sigma}_2(1+{y}\eta)}{[a\eta^2+(1+{y}\eta)^2]^2}\bigl[a\eta\eta_{;2}+(1+y\eta)(y_2\eta+y\eta_{;2}) \bigr],
\nonumber\\
{y}_{21}=\frac{\partial^2\,{y}}{\partial{\phi}_2\partial{\phi}_1}=\frac{\bar{\epsilon}^{(r)}({\sigma}_{2;1}\eta+\sigma_2\eta_{;1}) -{\sigma}_{1;1}(1+{y}\eta)-
\sigma_1(y_1\eta+y\eta_{;1})}{a\eta^2+(1+{y}\eta)^2}
\nonumber\\
-2\frac{\bar{\epsilon}^{(r)}{\sigma}_2\eta-{\sigma}_1(1+{y}\eta)}{[a\eta^2+(1+{y}\eta)^2]^2}\bigl[a\eta\eta_{;1}+(1+y\eta)(y_1\eta+y\eta_{;1}) \bigr],
\label{nw13.0zf3}
\end{eqnarray}
where $\eta_{;j}=\partial\eta/\partial{\phi}_j$ and ${\sigma}_{i;j}=\partial{\sigma}_i/\partial{\phi}_j,\,\,\,( i,j=1,2).$\\
It is important to note that the antisymmetry of the non-diagonal elements of the metric tensor arises at the stage of choosing a coordinate system and, accordingly, the orientation of the sub-manifold 
under consideration $\Xi^2_{\{{\bm\phi}\}}$.

Regarding the question of the symmetry of mixed second derivatives on any oriented manifolds, then, based on the basic requirement of mathematical analysis, the following identity must be satisfied at 
any point in four-dimensional Minkowski space (Schwartz’s theorem, see \cite{Rud}):
\begin{equation}
{y}_{12}=\frac{\partial^2\,{y}}{\partial{\phi}_1\partial{\phi}_2}={y}_{21}=\frac{\partial^2\,{y}}{\partial{\phi}_2\partial{\phi}_1},
 \label{q4.tz04}
\end{equation}
which is a necessary condition for a twice continuously differentiable function. In~the context of partial differential equations, it is called the Schwarz integrability~condition.

Using equations (\ref{nw3.02}), (\ref{q4.t04}) and (\ref{q4.tz04}), we can finally obtain the following fourth-degree algebraic equation for the asymmetric element of the metric tensor $g^{ 12}=-g^{21}=y$:
\begin{eqnarray}
\sum_{n=0}^4 A_n(\phi_1,\phi_2|u,s)y^n=0,
\label{nw3z.0zf3}
\end{eqnarray}
where the coefficients of the algebraic Equation $A_n(\phi_1,\phi_2|u,s)$ are defined by the expressions:
$$
A_0 =\varepsilon^{(r)}\varepsilon^{(i)}\bigl\{4\varepsilon^{(r)}\varepsilon^{(i)}b_1\phi_1 -4\varepsilon^{(r)}a^{(i)}_1b_1\phi_1^2\phi_2-2a^{(r)}_1\varepsilon^{(r)}\phi_1\phi_2(a^{(i)}_1+2b_1\phi_2)
+2a^{(i)}_1b_1\varepsilon^{(i)}\phi_2(\phi_1^2-\phi_2^2) \bigr\} 
$$
$$
 -[a_1^{(i)}]^2(\varepsilon^{(r)}\phi_1^2+\varepsilon^{(i)}\phi_2^2) -[a_1^{(r)}]^2(\varepsilon^{(i)}\phi_1^2+\varepsilon^{(r)}\phi_2^2)+2\varepsilon^{(i)}a^{(r)}\bigl[\varepsilon^{(r)}+a^{(i)}\phi_1\phi_2-b_1\phi_1(\phi_1^2-\phi_2^2)\bigr]
$$
$$
-b^2_1\bigl[4\varepsilon^{(r)}\phi_1^2\phi_2^2+\varepsilon^{(i)}(\phi_1^2-\phi_2^2)^2\bigr]\bigl\},\,\,\,  A_1=-\epsilon^{(r)}\epsilon^{(i)}\bigl(\epsilon^{(r)}+\epsilon^{(i)}\bigr)\bigl(a^{(i)}+2b_1\phi_2\bigr),\,\,\, A_2= 2\bigl\{4\varepsilon^{(r)}\times
$$
$$
a_1^{(i)}b_1\phi_1^2\phi_2+[a_1^{(i)}]^2(\varepsilon^{(r)}\phi_1^2+\varepsilon^{(i)}\phi_2^2)+[a_1^{(r)}]^2(\varepsilon^{(i)}\phi_1^2+\varepsilon^{(r)}\phi_2^2)+b_1^2\bigl[4\varepsilon^{(r)}\phi_1^2\phi_2^2+\varepsilon^{(i)}(\phi_1^2-\phi_2^2)^2\bigr]
$$
$$
+\,2b_1\varepsilon^{(i)}\bigl[6\varepsilon^{(r)}\phi_1\,-\,a_1^{(i)}\phi_2(\phi_1^2\,-\,\phi_2^2)\bigr]\,+\,2a^{(r)}\bigl[\varepsilon^{(r)}\phi_1\phi_2(a^{(i)}_1+2b_1\phi_2)\,+\,\varepsilon^{(i)}(3\varepsilon^{(r)}\,-\,a^{(i)}\phi_1\phi_2\,
$$
$$
+\,b_1\phi_1(\phi_1^2-\phi_2^2)\bigr]\bigr\},\quad A_3=-4\bigl(\varepsilon^{(r)}+\varepsilon^{(i)}\bigr)\bigl(a^{(i)}+2b_1\phi_2\bigr),\quad A_4=16\bigl(a^{(r)}_1+2b_1\phi_1\bigr).
$$
\begin{figure} 
\centering
\includegraphics[width=70mm]{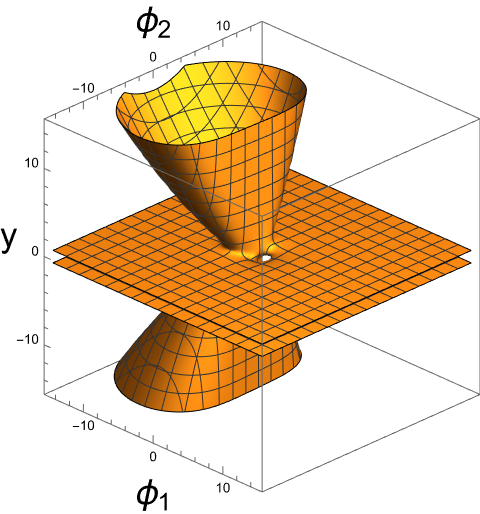}
\qquad
\includegraphics[width=70mm]{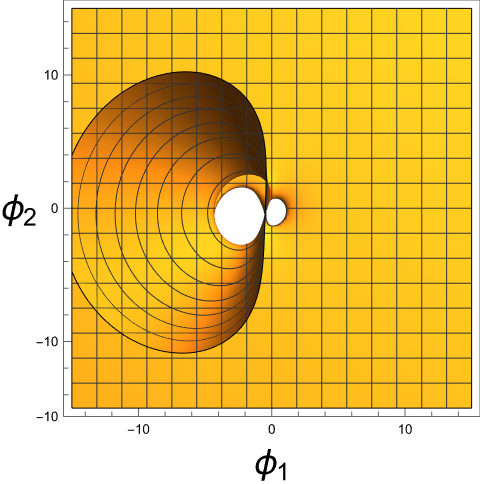}
\caption{\emph{The left figure is a three-dimensional plot of  the metric tensor element $g^{12}(\phi_1,\phi_2)$ in the limit $s\to\infty$, calculated in nucleon ground state for values of $\varepsilon^{(r)}=\varepsilon^{(i)}=1.5$ and $a^{(r)}=0,
\,\,a^{(i)}=b_1=1$. The right figure shows a two-dimensional projection of the sub-manifold $\Xi_{\{{\bm\phi}\}}^2$, where it is clearly seen that the its has a singularity with a Betti number of 2.}
\label{overflow12}}
\end{figure}
As can be seen, at each point of Minkowski space-time  $ u\in \mathbb{R}^4$ the coefficients of the Equation (\ref{nw3z.0zf3})  are functions of two coordinates $(\phi_1, \phi_2)$, which are 
the coordinates of the tangent bundle of the two-dimensional sub-manifold  $\Xi_{\{{\bm\phi}\}}^2$.  Note that the sub-manifold $\Xi_{\{{\bm\phi}\}}^2$ generates an algebraic Equation (\ref{nw3z.0zf3})
 in the form of a two-dimensional quantized space (again due to the dependence on the function $a_k( u,s)$ see (\ref{q3.02})). In particular, depending on the value of the fluctuation powers 
 $\bar{\varepsilon}^{(r)}$ and $\bar{\varepsilon}^{(i)}$ in the continuum set of points defined by the coordinates $(\phi_1,\phi_2)$, the solution to the algebraic equation can be complex. In this case,
it is  necessary to cut out such regions and leave only  the regions in which the algebraic Equation has real solutions. As a result of this procedure, the resulting sub-manifold will have topological features 
characterized by the Betti number $n\leq4$ (see Fig.s 3 and 4), where 4 is the number of complex solutions of the algebraic Equation (\ref{nw3z.0zf3}).

Note that a similar proof can be carried out in the case of Equations of distributions; $\mathcal{P}_{\bm\varphi}({\bm\varphi},s|v)$ and $\mathcal{P}_{\bm\theta}({\bm\theta},s|w)$.
It is obvious that the generating additional two-dimensional sub-manifoldes $\Xi^2_{\{{\bm\varphi}\}}$ and $\Xi^2_{\{{\bm\theta}\}}$ will have similar geometric and topological features 
as the sub-manifold $\Xi^2_{\{{\bm\phi}\}}$.
  \end{proof}


\section{C\lowercase{onstruction of a measure of the functional sub-space }} 
To calculate the mathematical expectation of different parameters of a  dynamical system, we need to construct measures of three functional sub-spaces ${\Xi}_{\{{\bm\phi}(s)\}}$,\, 
${\Xi}_{\{{\bm\varphi}(s)\}}$ and ${\Xi}_{\{{\bm\theta}(s)\}}$. Since all these noted sub-spaces are similar in their geometric and topological properties, below we will study only one 
of these sub-spaces and construct its measure.

 Let the probability distribution in each point of the Minkowski space-time $u\in \mathbb{R}^4$ satisfy the following limiting condition:
 \begin{equation}
\lim_{s\to s'} \mathcal{P}_{\bm\phi}({\bm{\phi}},s|{\bm{\phi}'},s')=\delta({\bm{\phi}}-{\bm{\phi}'}),\qquad  s=s'+\Delta s,
 \label{q6.01}
 \end{equation}
 
Taking into account (\ref{q6.01}) for small intervals of events, that is, for $\Delta s=s-s'\ll 1$, we can present the solution to Equation (\ref{q4.01})-(\ref{q4.02}) in the form (see also \cite{Ash1}):
\begin{eqnarray}
\label{6q.02}
P_{{\bm\phi}}({\bm\phi},s;u|{\bm\phi}',s')=\frac{1}{2\pi\sqrt{|\det{\bar{\bm\varepsilon}}|}\,\Delta s}\times
\qquad\qquad\qquad\qquad\qquad
\nonumber\\
\exp\Biggl\{-\frac{\bigl[{\bm\phi}-{\bm\phi}'-{\bm{\sigma}}({\bm\phi},s|u)\Delta s\bigr]^T{\bar{\bm\varepsilon}}^{-1}
\bigl[{\bm\phi}-{\bm\phi}'-{\bm{\sigma}}({\bm\phi},s|u)\Delta s\bigr]}{2\Delta s}\Biggr\},\qquad
\end{eqnarray}
where $\bar{\bm\varepsilon}$ is the second-rank matrix with elements $\varepsilon_{11}=\bar{\varepsilon}^{(r)},\,\,\varepsilon_{22}
=\bar{\varepsilon}^{(i)}$ and $\varepsilon_{12}=\varepsilon_{21}=0$, while $[\cdot\cdot\cdot]^T$ denotes a vector~transposition. 

Additionally, in representation (\ref{6q.02}), a two-dimensional vector ${\bm{\sigma}}(\bm{\phi},s|u)$ is defined as:
\begin{equation}
{\bm{\sigma}}({\bm\phi},s|u)= \Biggl\{
\begin{array}{ll} 
 \sigma_1({\bm\phi},s|u)=a_1^{(r)}\phi_1\,-a_1^{(i)}\phi_2+\,b_1[\phi^2_1-\phi_2^2],
\\
\sigma_2({\bm\phi},s|u)\,=a_1^{(i)}\phi_1+a_1^{(r)}\phi_2+2b_1\phi_1\phi_2,
\end{array}
\label{q6.03}
\end{equation}
where the functions $\sigma_1({\bm\phi},s|u)$ and $\sigma_2({\bm\phi},s|u)$ implicitly depend on the event interval $``s"$ and parametrically on the points Minkowski space $u\in \mathbb{R}^4$.
 
As can be seen from the expression (\ref{6q.02}), the~evolution of the system in the functional space ${\Xi}_{\{{\bm\phi}(s)\}}$ is characterized by a regular shift 
with  a speed ${\bm{\sigma}}({\bm{\phi}},s|u)$ against the background of Gaussian fluctuations with the diffusion matrix $\varepsilon_{ij}$.
 Concerning to the trajectory ${\bm{\phi}}(s)$ in the functional space  ${\Xi}_{\{{\bm\phi}(s)\}}$, it is determined by the following Equations (see ~\cite{Gard}):
\begin{equation}
{\bm{\phi}}(s)= \Biggl\{
\begin{array}{ll} 
\phi_1(s+\Delta s)= \phi_1(s)+\sigma_1({\bm\phi},s|u)\Delta s +(\Delta s)^{1/2}\,f^{(r)}(s),
\\
\phi_2(s+\Delta s)= \phi_2(s)+\sigma_2({\bm\phi},s|u)\Delta s + (\Delta s)^{1/2}\,f^{(i)}(s).
\end{array}
\label{q6.03}
\end{equation}
In order not to complicate the writing of formulas, we do not write the parametric dependence of the functions $\phi_1$ and $\phi_2$ on the variable $``u$''.

As can be seen from (\ref{q6.03}), the trajectory ${\bm{\phi}(s)}$ is continuous everywhere, since ${\bm{\phi}}(s+\Delta s)\bigl |_{\Delta s\to 0}={\bm{\phi}(s)}$, but 
nevertheless it is non-differentiable everywhere due to the presence of the term $(\Delta s)^{1/2}$.  If the interval of events is represented as $\Delta s= s/N$, 
where $N\to\infty$, then expression (\ref{6q.02}) can be interpreted as the probability of transition from the vector field ${\bm\phi}_l (s)$ to the vector field 
${\bm\phi}_{l+1}(s)$,  during of the interval $\Delta s$ within the Brownian motion model.

Finally, we can define the measure of the function space ${\Xi}_{\{{\bm\phi}(s)\}}$, which we will conventionally call the Fokker–Planck measure:
\begin{eqnarray}
\label{q6.04}
D\mu(\bm{\phi})=d\mu(\bm{\phi}_0)\lim_{N\to\,\infty}\Biggl\{
\biggl(\frac{1}{2\pi} \frac{N/s}{\sqrt{\bar{\varepsilon}^{(r)}\bar{\varepsilon}^{(i)}}}\,\biggr)^N\prod_{l=\,0}^N
d{\phi}_{1(l+1)}d{\phi}_{2(l+1)}\exp\Biggl[-\frac{N/s}{2{\bar\varepsilon}^{(r)}}\biggl({\phi}_{1(l+1)}
\nonumber\\
-{\phi}_{1(l)}-\sigma_{1(l+1)}\frac{s_{l+1}}{N}\biggr)^2-\frac{N/s}{2\bar{\varepsilon}^{(i)}}\biggl({\phi}_{2(l+1)}-
{\phi}_{2(l)}-\sigma_{2(l+1)}\frac{s_{l+1}}{N}\biggr)^2\Biggr]\Biggr\},
\end{eqnarray}
where $d\mu({\bm\phi}_0)=\delta(\phi_{1}-\phi_{1(0)})\delta(\phi_{2}-\phi_{2(0)})d\phi_{1}d\phi_{2}$
denotes the measure of the initial distribution,  in~addition, in the representation (\ref{q6.04}) the following notations are made:
$$
\phi_{1(l)}=\phi_1(s_l),\quad \phi_{2(l)}=\phi_2(s_l),\quad \sigma_{1(l)}=\sigma_1(\phi_{1(l)},\phi_{2(l)},s_l),
\quad \sigma_{2(l)}=\sigma_2(\phi_{1(l)},\phi_{2(l)},s_l).
$$

Thus, we have constructed a measure (\ref{q6.04}) of the functional sub-space ${\Xi}_{\{{\bm\phi}(s)\}}$, which is necessary for further analytical constructions of the theory.
In a similar way, we can construct expressions for the measures of the sub-spaces  ${\Xi}_{\{{\bm\varphi}(s)\}}$ and ${\Xi}_{\{{\bm\theta}(s)\}}$, respectively.

\section{M\lowercase{athematical expectation of the nucleon wave function }} 
To calculate the mathematical expectation of the full wave function, we first need to average the complex probabilistic process (\ref{q3.01}) over the functional sub-space ${\Xi}_{\{{\bm\chi}(s)\}}$.

{\bf Definition 1.} \emph{Let us call the mathematical expectation of a complex probabilistic process taking into account the influence of colored gluon fields  the following integral representation:}
\begin{equation}
\overline{\Psi}_{k}(\xi)=\mathbb{E}[\breve{\Psi}_k(\xi|{\bm\chi})]=\frac{\Psi_{k}(\xi)}{\alpha_k(s,\xi)}{\int\cdot\cdot\cdot\int}_{{\Xi}_{\{{\bm\chi}(s)\}}} D\mu(\bm\chi)
\exp\biggl( \int_0^{s}\Lambda_k(s';{\xi})ds'\biggr),
\label{q7.01}
\end{equation}
\emph{where  $\alpha_k(s,\xi)={\int\cdot\cdot\cdot\int}_{{{\Xi}_{\{{\bm\chi}(s)\}}}}D\mu(\bm\chi)=\int\int_{{\Xi}^2_{\{{\bm\chi}\}}}\mathcal{P}_{\bm\chi}({\bm\chi},s)d\chi_1d\chi_2$ is a normalizing constant}.\\
Recall that we consider that $\breve{\Psi}_k\bigl(\xi|\lambda_1(s_\xi)\bigr)=\breve{\Psi}_k(\xi|{\bm\chi})$ (see expression (\ref{q2.W03})).

Using the generalized Feynman-Kac theorem \cite{Ash2,Ash3} taking into account (\ref{q3.01}), we can integrate the functional integral into (\ref{q7.01}) and find the following two-dimensional 
integral representation of the mathematical expectation of the wave function:
\begin{equation}
\overline{\Psi}_1(u)=\mathbb{E}[\breve{\Psi}_1(u|\bm\phi)]=\frac{\Psi_{1}(u)}{\alpha_1(s,u)}\int\int_{{\Xi}^2_{\{\bm\phi\}}} Q_{\bm\phi}({\bm\phi},s|u)d\phi_1d\phi_2,
\label{q7.02}
\end{equation}
where the integrand function $ Q_{\bm\phi}({\bm\phi},s|u)$ is the solution of the following complex PDE:
\begin{equation}
\frac{\partial{Q}_{\bm\phi}}{\partial s}=\bigl\{\widehat{\mathcal{L}}^{(0)}_{\bm\phi}({\bm\phi},s|u)+\phi_1+i\phi_2\bigr\}{Q}_{\bm\phi}.
\label{q7.03}
\end{equation}
Let us represent the solution to the Equation (\ref{q7.03}) as the sum of the real and imaginary parts:
\begin{equation}
Q_{\bm\phi}({\bm\phi},s|u) =Q^{(r)}_{\bm\phi}({\bm\phi},s|u)+iQ^{(i)}_{\bm\phi}({\bm\phi},s|u).
\label{q7.04}
\end{equation}
Substituting (\ref{q7.04}) into the Equation (\ref{q7.03}), we get two real coupled PDEs:
\begin{equation}
 \Biggl\{
\begin{array}{ll} 
\partial_s{Q}_{\bm\phi}^{(r)}=\bigl\{\widehat{\mathcal{L}}^{(0)}_{\bm\phi}+\phi_1\bigr\}{Q}_{\bm\phi}^{(r)}-\phi_2Q_{\bm\phi}^{(i)},
\\
\partial_s{Q}_{\bm\phi}^{(i)}=\bigl\{\widehat{\mathcal{L}}^{(0)}_{\bm\phi}\,+\phi_1\bigr\}{Q}_{\bm\phi}^{(i)}+\phi_2Q_{\bm\phi}^{(r)}.
\end{array}
 \label{q7.05}
\end{equation}
Similar to the definition (\ref{q7.01}), one can construct mathematical expectations of complex probabilistic processes $\breve{\Psi}_2(v|{\bm\varphi})$ and $\breve{\Psi}_3(w|{\bm\theta})$.

In particular, averaging the wave function $\breve{\Psi}_2(v|{\bm\varphi})$ we find the following expression:
\begin{equation}
\overline{\Psi}_2(v)=\mathbb{E}[\breve{\Psi}_2(v|\bm\varphi)]=\frac{\Psi_{2}(v)}{\alpha_2(s,v)}\int\int_{{\Xi}^2_{\{\bm\varphi\}}} \bigl[Q^{(r)}_{\bm\varphi}({\bm\varphi},s|v)
+iQ^{(i)}_{\bm\varphi}({\bm\varphi},s|v)\bigr]d\varphi_1d\varphi_2,
\label{q7.06}
\end{equation}
where $\alpha_2(s,v)={\int\cdot\cdot\cdot\int}_{{{\Xi}_{\{{\bm\varphi}(s)\}}}}D\mu(\bm\varphi)=\int\int_{{\Xi}^2_{\{{\bm\varphi}\}}}\mathcal{P}_{\bm\varphi}({\bm\varphi},s)d\varphi_1d\varphi_2$
and, accordingly, $Q^{(r)}_{\bm\varphi}({\bm\varphi},s|v)$ and $Q^{(i)}_{\bm\varphi}({ \bm\varphi},s|v)$ are solutions to the following PDEs' system:
\begin{equation}
 \Biggl\{
\begin{array}{ll} 
\partial_s{Q}_{\bm\varphi}^{(r)}=\bigl\{\widehat{\mathcal{L}}^{(0)}_{\bm\varphi}+\varphi_1\bigr\}{Q}_{\bm\varphi}^{(r)}-\varphi_2Q_{\bm\varphi}^{(i)},
\\
\partial_s{Q}_{\bm\varphi}^{(i)}=\bigl\{\widehat{\mathcal{L}}^{(0)}_{\bm\varphi}\,+\varphi_1\bigr\}{Q}_{\bm\varphi}^{(i)}+\varphi_2Q_{\bm\varphi}^{(r)}.
\end{array}
 \label{q7.d05}
\end{equation}
The procedure for functional averaging of the wave function $\breve{\Psi}_3(w|{\bm\theta})$ leads to the following result:
\begin{equation}
\overline{\Psi}_3(w)=\mathbb{E}[\breve{\Psi}_3(w|\bm\theta)]=\frac{\Psi_{3}(w)}{\alpha_3(s,w)}\int\int_{{\Xi}^2_{\{\bm\theta\}}} \bigl[Q^{(r)}_{\bm\theta}({\bm\theta},s|w)
+iQ^{(i)}_{\bm\theta}({\bm\theta},s|w)\bigr]d\theta_1d\theta_2,
\label{q7.07}
\end{equation}
where $\alpha_3(s,w)={\int\cdot\cdot\cdot\int}_{{{\Xi}_{\{{\bm\theta}(s)\}}}}D\mu(\bm\theta)=\int\int_{{\Xi}^2_{\{{\bm\theta}\}}}\mathcal{P}_{\bm\theta}({\bm\theta},s)d\theta_1d\theta_2$, in addition, the functions
$Q^{(r)}_{\bm\theta}({\bm\theta},s|w)$ and $Q^{(i)}_{\bm\theta}({\bm\theta},s|w)$ are  solutions to the following PDEs' system:
\begin{equation}
 \Biggl\{
\begin{array}{ll} 
\partial_s{Q}_{\bm\theta}^{(r)}=\bigl\{\widehat{\mathcal{L}}^{(0)}_{\bm\theta}+\theta_1\bigr\}{Q}_{\bm\theta}^{(r)}-\theta_2Q_{\bm\theta}^{(i)},
\\
\partial_s{Q}_{\bm\theta}^{(i)}=\bigl\{\widehat{\mathcal{L}}^{(0)}_{\bm\theta}\,+\theta_1\bigr\}{Q}_{\bm\theta}^{(i)}+\theta_2Q_{\bm\theta}^{(r)}.
\end{array}
 \label{q7.w05}
\end{equation}

Now, regarding the functions $Q_{\bm\chi}^{(\upsilon)}({\bm\chi},s|\xi),$ where $(\upsilon=r,i),\,{\bm\chi=({\bm\phi,{\bm\varphi},{\bm\theta}})}$ and $\xi=(u,v,w)$, giving them the meaning of density probabilities, 
we can normalize them:
\begin{equation}
\bar{Q}_{\bm\chi}^{(\upsilon)}({\bm\chi},s|\xi)=\alpha_{\bm\chi}^{-1}(\xi;s){Q}_{\bm\chi}^{(\upsilon)}({\bm\chi},s|\xi), 
 \label{q7.08}
\end{equation}
where
$
\alpha_{\bm\chi}(\xi;s)=\int\int_{{\Xi}^2_{\{\bm\chi\}}}\sum_{\upsilon=r,i}Q_{\bm\chi}^{(\upsilon)}({\bm\chi},s|\xi)d\chi_1d\chi_2.
$

Obviously, for the normalized probability distributions the following condition will occur:
$$
 \int\int_{{\Xi}^2_{\{{\bm\chi}\}}}\sum_{\upsilon=r,i}\bar{Q}_{\bm\chi}^{(\upsilon)}({\bm\chi},s|\xi)d\chi_1d\chi_2=1.
$$ 
Finally, taking into account the obtained results (\ref{q2.W03}), (\ref{q7.02}), (\ref{q7.06}) and (\ref{q7.07}), we write down the mathematical expectation of the wave nucleon functions taking into 
account the continuous gluon exchange between quarks and the influence of the sea of colored quarks:
\begin{equation}
\overline{\Psi}(u,v,w)=\mathbb{E}[\breve{\Psi}(u,v,w|\{\lambda\})]=\overline{\Psi}_1(u)\overline{\Psi}_2(v)\overline{\Psi}_3(w).
 \label{q7.0rt7}
\end{equation}
By carrying out this type of normalization, we actually take into account two-quark interactions in a three-quark system, which is a simplification of the real problem.  To take into account three-quark 
interactions that preserve the color of the nucleon, it is necessary to carry out more complex three-color synchronization in the dynamical system, which will be equivalent to taking into account 
three-quark interactions (see {\bf{Appendix}} for details).

In the end, we note that the mathematical algorithm for the numerical study of a complex PDE of type (\ref{q7.03}) has been studied in detail in the works of the authors \cite{Ash,Ash1}.

\section{M\lowercase{athimatical expectation of nucleon radius and mass }}
The average radius and average mass of a nucleon are formed as mathematical expectations of the corresponding quantities in the ground state of the nucleon.

{\bf{Definition 2.}} \emph{The radius of the nucleon will be called the square root of the average value of the square of the radii of quark displacements, calculated in the ground state of the nucleon:}
\begin{eqnarray}
\mathsf{R}_{nuc}\bigl(\varepsilon^{(r)}_1,\cdot\cdot\cdot,\varepsilon^{(i)}_3\bigr)=\lim_{s\to\,\infty}\sqrt{\int\cdot\cdot\cdot\int({\bf{v}}^2+{\bf{w}}^2)\varrho(u,v,w)d^4{v}d^4{ w}}, 
 \label{q10.01}
\end{eqnarray}
\emph{where  ${\bf{v,w}}\in \mathbb{R}^3$  denote  the spatial three-dimensional vectors,   $\varrho_0(u,v,w)=|\overline{\Psi}_0(u,v,w)|^2$ is the probability density  
and  $\overline{\Psi}_0(u,v,w)$ denotes the mathematical expectation of the wave function of the nucleon in the ground state}.\\
The fact that there is no integration over the coordinates of the four-dimensional vector $u$ is due to the fact that the system performs translational motion along these coordinates.
The expression (\ref{q10.01}) can be written explicitly:
\begin{eqnarray}
\mathsf{R}_{nuc}\bigl(\varepsilon^{(r)}_1,\cdot\cdot\cdot,\varepsilon^{(i)}_3\bigr)=\qquad\qquad\qquad \qquad\qquad\qquad\qquad\qquad\qquad\qquad\qquad\qquad\qquad\quad
\nonumber\\
\lim_{s\to\,\infty}\sqrt{\int\cdot\cdot\cdot\int e^{-\frac{\Omega_0}{3}\bigl(v_0^2+{\bf v}^2+w_0^2+{\bf w}^2\bigr)}({\bf{v}}^2+{\bf{w}}^2)
\frac{\varrho_0(v,w,s)d^4vd^4w}{\alpha^2_1(u,s)\alpha^2_2(v,s)\alpha^2_3(w,s)}}, 
 \label{q10.a01}
\end{eqnarray}
where the function of the probability density $\varrho_0(v,w,s)$  is defined as follows:
\begin{eqnarray}
\varrho_0(v,w,s)=\biggl| \int\int_{{\Xi}^2_{\{\bm\phi\}}}\bigl[Q^{(r)}_{\bm\phi}({\bm\phi},s|u)
+iQ^{(i)}_{\bm\phi}({\bm\phi},s|u)\bigr]d\phi_1d\phi_2 \times\int\int_{{\Xi}^2_{\{\bm\varphi\}}} \bigl[Q^{(r)}_{\bm\varphi}({\bm\varphi},s|v)\qquad
\nonumber\\
+\,iQ^{(i)}_{\bm\varphi}({\bm\varphi},s|v)\bigr]d\varphi_1d\varphi_2  \times
\int\int_{{\Xi}^2_{\{\bm\theta\}}} \bigl[Q^{(r)}_{\bm\theta}({\bm\theta},s|w)+iQ^{(i)}_{\bm\theta}({\bm\theta},s|w)\bigr]d\theta_1d\theta_2\biggr|^2.\,\,\,\,\,
 \label{q10.a02}
\end{eqnarray}
 
Finally, we can calculate an important parameter of the nucleon - its mass.

{\bf Definition 3.} \emph{The mathematical expectation of the nucleon mass in the ground state will be the following expression:}
\begin{eqnarray}
m_{nuc}\bigl(\varepsilon^{(r)}_1,\cdot\cdot\cdot,\varepsilon^{(i)}_3\bigr)=\lim_{s\to\,\infty}\sqrt{\int\cdot\cdot\cdot\int\Bigl|\overline{\Psi}^\ast(u,v,w)\frac{\partial^2}{\partial t^2}\overline{\Psi}(u,v,w) d^4vd^4w\Bigr|}.
 \label{q10.a03}
\end{eqnarray}
Having calculated (\ref{q10.a03}), we obtain the following expression for the nucleon mass:
\begin{eqnarray}
m_{nuc}\bigl(\varepsilon^{(r)}_1,\cdot\cdot\cdot,\varepsilon^{(i)}_3\bigr)=\qquad\qquad\qquad\qquad\qquad\qquad\qquad\qquad\qquad\qquad\qquad\qquad\qquad\,\,
\nonumber\\
m_0\,\lim_{s\to\,\infty} \sqrt{\int\cdot\cdot\cdot\int e^{-\frac{\Omega_0}{3}\bigl(v_0^2+{\bf v}^2+w_0^2+{\bf w}^2\bigr)}\frac{\varrho_0(v,w,s)d^4vd^4w}{\alpha^2_1(u,s)\alpha^2_2(v,s)\alpha^2_3(w,s)} }.
 \label{q10.a04}
\end{eqnarray}

After we have determined the radius and mass of the nucleon through the fluctuation powers of the gluon fields $(\varepsilon_1^{(r)},...,\varepsilon_3^{ (i)})$, a natural question arises, namely: can we find 
these constants, knowing only the masses of the nucleon and quarks, respectively? Based on the fact that two of the three quarks in a nucleon are the same, and the fluctuation powers of the gluon and antigluon
fields are equal, we can conclude that there are only two independent constants that characterize six-color gluon fields. The latter means that if the mass of a nucleon in a free state is known, then in addition to equation 
(\ref{q10.a04}) another equation is needed to completely determine these constants. The missing second equation in this case may be the total energy of the colored gluon fields, which is equivalent to the effective 
mass of the sea of quarks $m_{qs}=m_{nuc} -m_0\simeq0.99\cdot m_{nuc }$. This energy is determined by the probability distributions on the corresponding two-dimensional sub-manifolds and can be represented 
as a summ of three terms (see expression (\ref{Azk9.k0w})):
\begin{eqnarray}
\lim_{s\to\,\infty}\sum_{\bm\chi=\bm\phi,\bm\varphi,\bm\theta}\,\sum_{\upsilon=i,r}\int\int_{\Xi^2_{\{{\bm\chi}\}}}\bar{Q}^{(\upsilon)}_{\bm\chi}({\bm\phi},s|\xi)d\chi_1d\chi_2={m_{qs}}/{m_{nuc}}\simeq0.99.
 \label{q10.a05}
\end{eqnarray}

Thus, we have obtained two equations (\ref{q10.a04}) and (\ref{q10.a05}), which allow us to uniquely determine two constants characterizing the powers of color gluon fluctuations in a free nucleon, when  
nucleon is in the \emph{ground state}.

\section{C\lowercase{onclusion}}
Although it has long been known that nucleons consist of valence quarks and a quark-antiquark sea, and although we know the rules of local gauge invariance with respect to the $SU(3)$ symmetry group, 
it can be argued that we are still far from a satisfactory understanding of the problem \cite{Antony}.  This is largely due to the problem of the strong nonlinearity of QCD, due to which perturbative methods 
are often inapplicable. In this regard, new, especially non-perturbative theoretical studies on this topic remain highly relevant and continue to serve as a source of ideas for new experiments in hadron physics.

The model of the nucleon as a three-quark relativistic system with interaction potentials of a four-dimensional harmonic oscillator, proposed by Feynman et al. \cite{Fey} turned out to be very useful for studying 
the structure and properties of the internal motion of the nucleon. However, as further experimental studies have shown, the rest mass of a three-quark system is only a percentage of the total rest mass of the nucleon. 
The missing main mass of the nucleon turns out to be due to the QCD binding energy, which arises as a result of the breaking of QCD chiral symmetry. In other words, intense interactions of valence quarks with colored 
gluon fields generate a sea of virtual quarks and antiquarks, which is ultimately recorded by measuring instruments as the rest mass of a nucleon.

To overcome this difficulty, we generalized the  relativistic model \cite{Fey}, considering the problem of internal motion nucleon as a self-organization problem of a complex three-quark dynamical system 
in a colored sea of quark-antiquark. We formulated the mathematical problem within the framework of a complex probabilistic process satisfying an equation of the Langevin-Kline-Gordon-Fock type 
(see Eq.s (\ref{q2.01})-(\ref{q2.04})). Using this equation, we obtained a system of SDEs that describes the motion of a six-color gluon field under the influence of a valence three-quark system 
(see system of Eq.s (\ref{q3.0zt2})). Note that the stochastic extension of the Klein-Gordon-Fock equation does not allow including all eight  gluns in the quark-gluon  interactions scheme due to  the neglect 
of the spin part of the motion. In this representation, the contributions of two colorless gluons are not taken into account. It is obvious that a complete description of the problem can only be achieved by stochastic 
extension of the Yang-Mills equations within the framework of the gauge symmetry group $SU(3)$, similar to what was done in the work of the author \cite {ASG} for the gauge symmetry group $SU(2)\bigotimes  U(1)$.
Recall that such work would be very important for quantum field theory also because it would make it possible to theoretically substantiate the existence of eight-color gluons.

At the first stage, using the system of SDEs (\ref{q3.0zt2}) in the limit of statistical equilibrium, a second-order PDE of six variables was obtained (see equation (\ref{q4.01})-(\ref {q4.02})), which describes the 
evolution of the distribution of colored gluon fields taking into account the self-interaction of gluons and the influence of the system of valence quarks. Using the symmetry property of the Equation 
(\ref{q4.01})-(\ref{q4.02}), we showed that the solution can be represented in factorized form as the product of three integral-representations, where the integrand satisfies second-order PDEs in two variables.
Analysis of geometric and topological properties of generated subspaces $\Xi^2_{\{{\bm\chi}\}},\,\,({\bm\chi}={\bm\phi},{\bm\varphi},{\bm\theta})$ shows that in the general case they are described by 
non-commutative geometry with topological features characterized by the Betti number $n\leq4$.  Note that the additional six-dimensional sub-space $\Xi^6_{\{{\bm\vartheta}\}}$ in this case is 
represented as a decomposition of direct products of two-dimensional quantized sub-manifolds (see expression (\ref{q4.w03z1})).  Nevertheless, the most important role of the additional subspace 
is that when averaging random processes over it that describe the joint system ``three valence quarks + sea of quarks'' and when constructing the mathematical expectation of the nucleon wave function
in Minkowski space, chiral symmetry is spontaneously violated. This can be easily verified by comparing the resulting representation for the mathematical expectation of the nucleon wave function (\ref{q7.0rt7}) 
with the wave function of a three-quark dynamical system, where quarks interact with each other through four-dimensional harmonic potentials (\ref{q2.03}). Figuratively speaking, the solution (\ref{q2.03}) 
describes the wave function in a potential with one minimum, and (\ref{q7.0rt7}) characterizes the wave state of the system with two or more local minima with spontaneous symmetry breaking. 

 At the second stage, the Eq.s for the distribution of gluon fields are derived, which is equivalent to the quark-antiquark sea generated by them, taking into account their intense exchange between quarks. 
As shown, in this case, the distributions of gluons and anti-gluons of the same color are related and specified using PDEs systems (see {\bf Appendix}, Eq.s  (\ref{qt4.0o2w})-(\ref{Az9.k0w})). By relating 
the distributions of gluon fields through joint normalization, we synchronize the colors of the quarks so that when they are mixed, the white color of the nucleon should  be preserved. Using these distributions, 
we constructed the mathematical expectation of the complete wave function of the nucleon (see equation (\ref{q7.0rt7}), as well as equations (\ref{q7.02}), (\ref{q7.06}) and (\ref{q7.07})), which, obviously, 
are no longer unitary and orthogonal.

Recall that if we find with a given accuracy the powers of gluon fluctuations $\varepsilon^{(\upsilon)}_k\,\,(\upsilon=i,r;\,\,k=1,2,3) $ for a nucleon in free state, then changes in these constants 
will obviously be directly related to the environment in which the nucleon is located. Nucleons located in dense and superdense stellar formations will have completely other parameters and 
structure due to changes in these constants.  In the future, we plan to numerically study the features of nucleons depending on these fluctuation powers. Recall that this will give us important information not 
only about nucleons, but also about macroscopic objects in which these particles are immersed. We emphasize that this kind of information about nucleons and their environment can only be obtained 
through the development of a nonperturbative theory.

Finally, it is important to note that all results obtained as the mathematical expectation of various nucleon parameters are formed in times $\Delta t\geq \hbar/2\Delta E=2,19\cdot10^{-24}\, sec$, where 
$\Delta E\approx m_{nuc}c_0^2$ is the energy of quark-antiquark sea  and $c_0$ is the speed of light in vacuum.

\section{A\lowercase{ppendix}}
As mentioned above, equation (\ref{q4.01})-(\ref{q4.02}) describes the distribution of gluon fields in the limit of statistical equilibrium, taking into account their interaction with the three-quark dynamical system,
as well as gluon-antigluon interactions (see terms $\sigma_2({\bm\phi},s|u)$, $\pi_2({\bm\varphi},s|u)$ and $\omega_2({\bm\theta},s|u)$ in the Equations (\ref{q4.02})-(\ref{q4.0z2})). This equation can be
 represented as:
\begin{eqnarray}
\frac{\partial\mathcal{P}}{\partial s} = \biggl\{\biggl(\bar{\varepsilon}_1^{(r)}\frac{\partial^2}{\partial \phi_1^2}+\bar{\varepsilon}_1^{(i)}\frac{\partial^2}{\partial \phi_2^2}\biggr)+
 \biggl(\bar{\varepsilon}_2^{(r)}\frac{\partial^2}{\partial \varphi_1^2}+\bar{\varepsilon}_2^{(i)}\frac{\partial^2}{\partial \varphi_2^2}\biggr)+ \biggl(\bar{\varepsilon}_3^{(r)}\frac{\partial^2}{\partial \theta_1^2}+
 \bar{\varepsilon}_3^{(i)}\frac{\partial^2}{\partial \theta_2^2}\biggr)+\quad
 \nonumber\\
\sum^2_{j=1}\Bigl[\frac{\partial}{\partial \phi_j} \sigma_j({\bm\phi},s |u)+ \frac{\partial}{\partial \varphi_j}\pi_j({\bm\varphi},s |v)+ \frac{\partial}{\partial \theta_j} \omega_j({\bm\theta},s |w)\Bigr]+
\mathcal{K}(\phi_1,\phi_2;\varphi_1,\varphi_2;\theta_1,\theta_2)\biggr\}\mathcal{P},
\label{q4t.02}
\end{eqnarray}
where $\mathcal{K}(\phi_1,\phi_2;\varphi_1,\varphi_2;\theta_1,\theta_2)=\mathcal{K}_{\bm\phi}+\mathcal{K}_{\bm\varphi}+\mathcal{K}_{\bm\theta}\equiv0$ is the color mixing member, in addition:
$$
\mathcal{K}_{\bm\phi}= \varphi_1+\varphi_2-\theta_1-\theta_2,\quad \mathcal{K}_{\bm\theta}=\phi_1+\phi_2-\varphi_1-\varphi_2,\quad \mathcal{K}_{\bm\varphi}=\theta_1+\theta_2-\phi_1-\phi_2.
$$
Substituting the solution (\ref{q4.03z}) into the equation (\ref{q4.01})-(\ref{q4.02}), we  find the following system of three loosely-coupled PDEs:\\
\begin{equation} 
\begin{cases}
\partial_s \mathcal{P}_{\bm\phi} =\widehat{\mathcal{L}}_{\bm\phi}({\bm\phi},s|u)\mathcal{P}_{\bm\phi}, \quad \widehat{\mathcal{L}}_{\bm\phi} = \bigl\{\bigl(\bar{\varepsilon}_1^{(r)}
\frac{\partial^2}{\partial \phi_1^2}+\bar{\varepsilon}_1^{(i)}\frac{\partial^2}{\partial \phi_2^2}\bigr)+\,\sum^2_{j=1}\frac{\partial}{\partial \phi_j} \sigma_j({\bm\phi},s |u)+\mathcal{K}_{\bm\phi}\bigr\},
\\
\partial_s \mathcal{P}_{\bm\varphi} =\widehat{\mathcal{L}}_{\bm\varphi}({\bm\varphi},s|v)\mathcal{P}_{\bm\varphi}, \quad \widehat{\mathcal{L}}_{\bm\varphi} = \bigl\{\bigl(\bar{\varepsilon}_2^{(r)}
\frac{\partial^2}{\partial \varphi_1^2}+\bar{\varepsilon}_2^{(i)}\frac{\partial^2}{\partial \varphi_2^2}\bigr)+\sum^2_{j=1}\frac{\partial}{\partial \varphi_j} \pi_j({\bm\varphi},s |v)+\mathcal{K}_{\bm\varphi}\bigr\},
\\
\partial_s \mathcal{P}_{\bm\theta}\, =\,\widehat{\mathcal{L}}_{\bm\theta}({\bm\theta},s|w)\mathcal{P}_{\bm\theta}, \quad \widehat{\mathcal{L}}_{\bm\theta} = \bigl\{\bigl(\bar{\varepsilon}_3^{(r)}
\frac{\partial^2}{\partial\theta_1^2}\,+\,\bar{\varepsilon}_3^{(i)}\frac{\partial^2}{\partial\theta_2^2}\bigr)+\,\sum^2_{j=1}\frac{\partial}{\partial \theta_j} \sigma_j({\bm\theta},s|w)+\mathcal{K}_{\bm\theta}\bigr\}. 
\label{Aq3.0zt2}
\end{cases}
\end{equation}
Now we can give the distributions $\mathcal{P}_{\bm\phi}({\bm\phi},s|u),\,\mathcal{P}_{\bm\varphi}({\bm\varphi},s|v)$ and $\mathcal{P}_{\bm\theta}({\bm\theta},s|w)$  the meaning of probability densities and normalize them:
\begin{eqnarray}
\mathcal{\bar{P}}_{\bm\phi}({\bm\phi},s|u,v,w)=C^{-1}(u,v,w,s)\mathcal{{P}}_{\bm\phi}({\bm\phi},s|u),
 \nonumber\\
\mathcal{\bar{P}}_{\bm\varphi}({\bm\varphi},s|u,v,w)=C^{-1}(u,v,w,s)\mathcal{{P}}_{\bm\varphi}({\bm\varphi},s|v),
 \nonumber\\
 \mathcal{\bar{P}}_{\bm\theta}({\bm\theta},s|u,v,w)\,=\,C^{-1}(u,v,w,s)\mathcal{{P}}_{\bm\theta}({\bm\theta},s|w), 
\label{qt4.0o2w}
\end{eqnarray}
where $C(u,v,w,s)$ is the normalization constant defined by the expression:
$$
C=\int\int_{\Xi^2_{\{{\bm\phi}\}}}\mathcal{{P}}_{\bm\phi}({\bm\phi},s|u)d\phi_1d\phi_2+\int\int_{\Xi^2_{\{{\bm\varphi}\}}}\mathcal{{P}}_{\bm\varphi}({\bm\varphi},s|v)d\varphi_1d\varphi_2
+\int\int_{\Xi^2_{\{{\bm\theta}\}}}\mathcal{{P}}_{\bm\theta}({\bm\theta},s|w)d\theta_1d\theta_2.
$$
Obviously, in this case the following equality holds:
\begin{eqnarray}
\int\int_{\Xi^2_{\{{\bm\phi}\}}}\mathcal{\bar{P}}_{\bm\phi}({\bm\phi},s|u,v,w)d\phi_1d\phi_2+\int\int_{\Xi^2_{\{{\bm\varphi}\}}}\mathcal{\bar{P}}_{\bm\varphi}({\bm\varphi},s|u,v,w)d\varphi_1d\varphi_2
\nonumber\\
+\int\int_{\Xi^2_{\{{\bm\theta}\}}}\mathcal{\bar{P}}_{\bm\theta}({\bm\theta},s|u,v,w)d\theta_1d\theta_2=1.
\label{At9.0o2w}
\end{eqnarray}

Now we can similarly construct a measure of the function space ${\Xi}_{\{{\bm \chi}(s )\}}$ and perform functional integration of the full wave function (see {\bf{Section}} VII and {\bf{Section}} VIII):
\begin{eqnarray}
\overline{\Psi}(u,v,w)\,=\,\mathbb{E}[\breve{\Psi}(u,v,w|\{\lambda\})]\,=\,\frac{{\Psi}_1(u){\Psi}_2(v){\Psi}_3(w)}{C(u,v,w,s)}\,\,\times\qquad\qquad\,\,\,\qquad
\nonumber\\
\prod_{\bm\chi=\bm\phi,\bm\varphi,\bm\theta}\,\int\int_{{\Xi}^2_{\{\bm\chi\}}} \bigl[{Q}^{(r)}_{\bm\chi}({\bm\chi},s|\xi)+i{Q}^{(i)}_{\bm\chi}({\bm\chi},s|\xi)\bigr]d\chi_1d\chi_2,\qquad \xi=u,v,w,
\label{A9.01}
\end{eqnarray}
where  the solutions ${Q}_{\bm\chi}^{(r)}$ and  ${Q}_{\bm\chi}^{(i)}$ satisfy the following system of PDEs:
\begin{equation}
 \Biggl\{
\begin{array}{ll} 
\partial_s{Q}_{\bm\chi}^{(r)}=\bigl\{\widehat{\mathcal{L}}_{\bm\chi}+\chi_1\bigr\}{Q}_{\bm\chi}^{(r)}-\chi_2{Q}_{\bm\chi}^{(i)},
\\
\partial_s{Q}_{\bm\chi}^{(i)}=\bigl\{\widehat{\mathcal{L}}_{\bm\chi}\,+\chi_1\bigr\}{Q}_{\bm\chi}^{(i)}+\chi_2{Q}_{\bm\chi}^{(r)}.
\end{array}
 \label{Aq7.05}
\end{equation}
As in the case of functions $\mathcal{{P}}_{\bm\phi},\,\mathcal{{P}}_{\bm\varphi}$ and $\mathcal{{P}}_{\bm\theta}$, we can give these solutions the meaning of probabilities density and normalize them.
In particular, one can write:
\begin{eqnarray}
\bar{Q}_{\bm\chi}^{(r)}(\bm\chi,s|u,v,w)=C_\ast^{-1}(u,v,w,s)Q_{\bm\chi}^{(r)}(\bm\chi,s|\xi),
\nonumber\\
\bar{Q}_{\bm\chi}^{(i)}(\bm\chi,s|u,v,w)=\,C_\ast^{-1}(u,v,w,s)Q_{\bm\chi}^{(i)}(\bm\chi,s|\xi), 
\label{At9.0ozw}
\end{eqnarray}
where $C_\ast(u,v,w,s)$ is the normalization constant, which is defined as follows:
\begin{eqnarray}
C_\ast(u,v,w,s)=\sum_{\bm\chi=\bm\phi,\bm\varphi,\bm\theta}\,\sum_{\upsilon=i,r}\int\int_{\Xi^2_{\{{\bm\chi}\}}}Q^{(\upsilon)}_{\bm\chi}({\bm\phi},s|\xi)d\chi_1d\chi_2. 
\label{Az9.k0w}
\end{eqnarray}
In addition, there will be conservation of total probability:
\begin{eqnarray}
\sum_{\bm\chi=\bm\phi,\bm\varphi,\bm\theta}\,\sum_{\upsilon=i,r}\int\int_{\Xi^2_{\{{\bm\chi}\}}}\bar{Q}^{(\upsilon)}_{\bm\chi}({\bm\phi},s|\xi)d\chi_1d\chi_2=1.
\label{Azk9.k0w}
\end{eqnarray}
Thus, we have combined and normalized the probability distributions of all quark colors, which allows us to be confident that we have synchronized the quark colors so that the nucleon color will be  invariant 
with respect to the internal motion of the nucleon. Obviously, such synchronization will mean taking into account three-quark interactions in the system.

Note that the choice of a model for connecting the probabilities ${Q}_{\bm\chi}^{(r)}(\bm\chi,s|u,v,w)$, ${Q}_{\bm\chi}^{(i)}(\bm\chi,s|u,v,w)$ or the parameter 
$\mathcal{K}(\phi_1,\phi_2;\varphi_1,\varphi_2;\theta_1,\theta_2)$ is not limited by any physical principle and therefore can be further refined taking into 
account the requirements of the considered problem. In the end, we note that if in the calculations replace the operator $\widehat{\mathcal{L}}_{\bm\chi}$ with 
$\widehat{\mathcal{L}}^{(0)}_{\bm\chi}=\widehat{\mathcal{L}}_{\bm\chi}\bigl|_{\mathcal{K}_{\bm\chi}=0}$ (see Eq.s (\ref{Aq3.0zt2})), then our consideration of the problem will have a two-particle approximation 
(see {\bf Section VIII}).

\section{A\lowercase{cknowledgments}}
Gevorkyan A.S. is grateful to grant N 21T-1B059 of the Science Committee of Armenia, which partially funded this work. The authors also thank Movsesyan K. A. for the numerical study and
visualization of calculations of two-dimensional sub-manifolds Fig.s 3-4.


\end{document}